\documentclass[runningheads]{rnl}
\usepackage{amssymb}
\usepackage[usenames,dvipsnames,svgnames,table]{xcolor}

\usepackage[latin1]{inputenc} 
\usepackage{amsmath,amsfonts,amssymb}

\usepackage{graphicx,epsf,subfigure} 
\newcommand{\greeksym}[1]{{\usefont{U}{psy}{m}{n}#1}}

\newcommand{\uDelta}{\mbox{\greeksym{D}}}

\def\BA{\begin{eqnarray}}
\def\EA{\end{eqnarray}}
\def\BAN{\begin{eqnarray*}}
\def\EAN{\end{eqnarray*}}
\let\NN=\nonumber
\def\sfrac#1#2{\mbox{$\frac{#1}{#2}$}}
\def\DDt{\sfrac{\rm d}{{\rm d} t}\,}

\def\CCB{\color{blue}}
\def\CCR{\color{red}}
\def\CCG{\color{OliveGreen}}

\def\UO{{\CCB {{\overline U}}_1}}
\def\BZ{{\CCB {B}_0}}
\def\DZ{{\CCB {D}_0}}
\def\FZ{{\CCB {F}_0}}
\def\AO{{\CCB {A}_1}}
\def\EO{{\CCB {E}_1}}
\def\CP{{\CCB {C}'_1}}
\def\GP{{\CCB {G}'_1}}

\def\UZ{{\CCR {{\overline U}}_0}}
\def\AZ{{\CCR {A}_0}}
\def\EZ{{\CCR {E}_0}}
\def\BO{{\CCR {B}_1}}
\def\FO{{\CCR {F}_1}}
\def\HP{{\CCR {H}'_1}}

\def\WZ{{\CCG {{\overline W}}_0}}
\def\CZ{{\CCG {C}_0}}
\def\GZ{{\CCG {G}_0}}
\def\HO{{\CCG {H}_1}}
\def\FP{{\CCG {F}'_1}}
\def\DP{{\CCG {D}'_1}}

\def\BP{B'_1}
\def\DO{D_1}

\def\HZ{H_0}
\def\WO{{{\overline W}}_1}
\def\CO{C_1}
\def\GO{G_1}
\def\AP{A'_1}
\def\EP{E'_1}

\begin{document}
\title*{Towards a model of large scale dynamics in transitional wall-bounded flows}
\titrecourt{Modelling wall-bounded flows}
\author{Paul Manneville}
\index{Manneville Paul}
\adresse{Laboratoire d'Hydrodynamique, UMR 7646, École Polytechnique, 91128 Palaiseau, France}
\email{paul.manneville@ladhyx.polytechnique.fr}

\maketitle 

\sloppy 

\begin{resumanglais}
A system of simplified equations is proposed to govern the feedback interactions of large-scale flows present in laminar-turbulent patterns of transitional wall-bounded flows with small-scale Reynolds stresses generated by the self-sustainment process of turbulence, itself modeled using an extension of Waleffe's approach~\cite{Wa97}.
This text is the English adaptation of~\cite{MaRNL}.
It is here complemented by an annex giving the detailed expression of the model alluded to above, absent from the French version due to space limitations.
\end{resumanglais}

\section{Context}
Flows controlled by viscous effects close to solid walls experience a subcritical transition to turbulence.
They stay linearly stable up to values of the Reynolds number $R$ high enough to enable nontrivial regimes arising from the nonlinearity of Navier--Stokes equations, in competition with the laminar regime.
Two examples with base flows stable against infinitesimal perturbations for all $R$ are representative of this situation: Hagen--Poiseuille flow in a cylindrical tube submitted to a pressure gradient, and plane Couette flow (PCF), the simple shear flow developing between two parallel plates in relative translation. 
Over a limited interval in $R$ called the {\it transitional regime\/}, this competition manifests itself in the form of pockets\\[0.5ex] 
\begin{minipage}{0.54\textwidth}
 filled with small-scale turbulent flow ({\it puffs\/}, {\it spots\/}) coexisting with laminar flow.
Specificities of this situation are review from an experimental, numerical, and theoretical point of vue in~\cite{Ma15}.
The transitional regime of PCF and other plane flows is characterized by the presence of wide oblique bands, alternatively laminar and turbulent~\cite{Petal03}, the understanding of which is made difficult by the existence of two spatiotemporal scales, one {\it large\/}, that of the laminar--turbulent pattern, the other {\it small\/}, internal to the regions agitated by {\it swift\/} turbulent eddies, as illustrated in the picture on the right.
\end{minipage}\hfill
\begin{minipage}{0.37\textwidth}
\null\hfill\includegraphics[width=\textwidth]{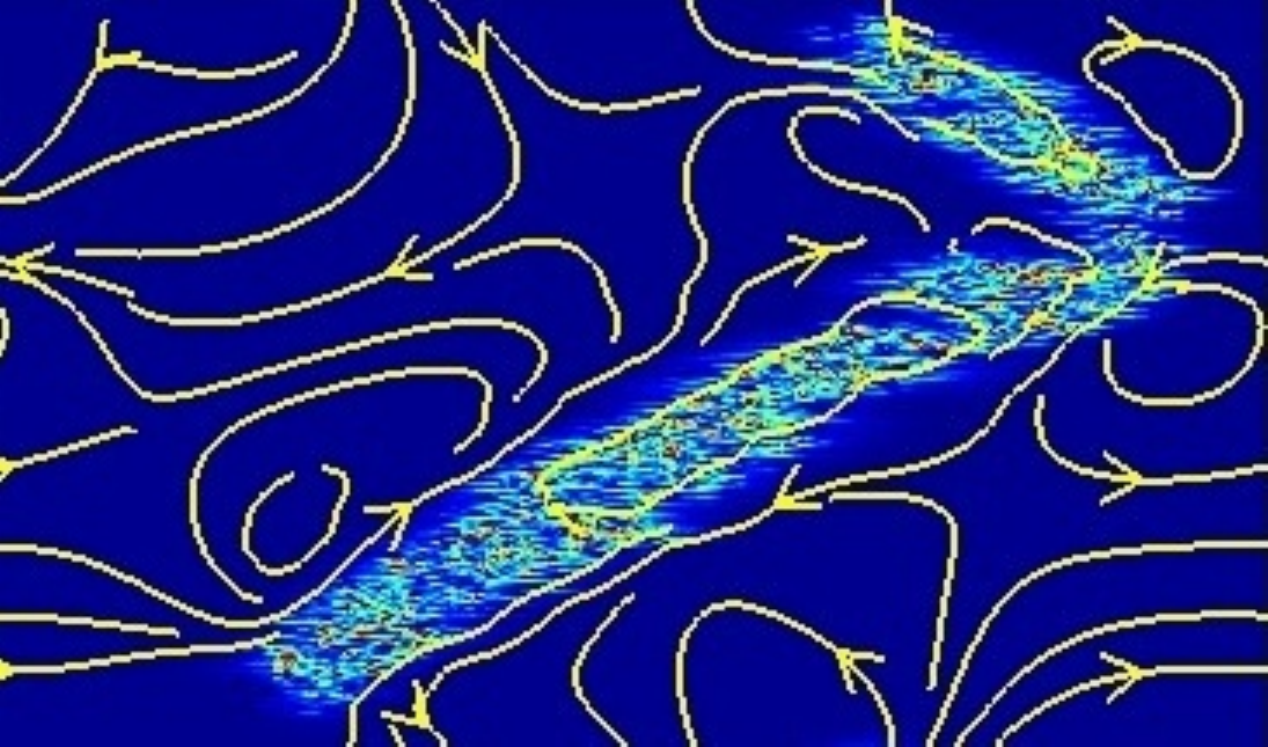}\\
\small Large-scale back-flow around turbulent band fragments in PCF; see~\cite{Ma1x} for details.
\end{minipage}\\[0.5ex]
\indent
The origin of turbulence at a local scale is believed to be well understood in terms of dynamics inside the {\it Minimal Flow Unit\/} (MFU)~\cite{JM91} where nontrivial solutions ``far'' from the laminar state are maintained by a {\it Self-Sustainment Process\/} (SSP)~\cite{Wa97}.
Within this framework, exploiting the local coherence of perturbations to the base flow, subsequent theoretical approaches have led to an interpretation of the transition using {\it deterministic chaos\/} theory typical of systems with a small number of degrees of freedom~\cite{Eetal08}, thus skipping all spatiotemporal aspects most appropriate for systems of experimental interest.
Local coherence can however serve as a starting point to a modeling strategy more adapted to the description of laminar--turbulent patterns present all along the transitional range~\cite[App. B]{Ma15}.
It indeed supports the idea that the dependence of the hydrodynamical fields in the wall-normal direction is more or less frozen while the dynamics in the complementary directions (one-dimensional in the pipe case, two-dimensional in the other cases) has regained most of its freedom.
This is what we shall consider in the following, restricting ourselves to the case of PCF though the approach has more general breadth.
In particular, we shall focus on the large-scale flows~\cite{BT07,Ma1x,DS13} that are more easily detected in the intervals between turbulent domains, within the framework --and using the tools-- of the general theory of {\it pattern formation\/} in far-from-equilibrium systems~\cite{Ho06}.

\section{Modeling}

PCF is the flow that develops between two parallel plates in relative translation in the streamwise direction denoted $x$; $y$ is the wall-normal direction; the spanwise direction $z$ completes the coordinate system.
The components of the perturbation to the base flow $u_{\rm b} = U y/h$ are $u$, $v$, and $w$;
$2U$ measures the relative speed of the plates and $2h$ the distance between them.
The Reynolds number is defined as  $R=U h / \nu$, where $\nu$ is the kinematic viscosity of the fluid.
We adopt a unit system based on $h$ for distances and $h/U$ for times.
In such a system, $R$ is numerically equal to $1/\nu$.
In order to simplify the treatment of pressure, we use a representation in which $v$ and $\zeta=\partial_z u-\partial_x w$ are the fundamental variables. 
The equations governing the perturbation to the base flow in this formulation are reviewed and discussed in the book by Schmid \& Henningson~\cite{SH01}, p.~155ff.

From previous approaches~\cite{Ma15} we shall keep in mind that the wall-normal dependence of the hydrodynamic fields is adequately described at lowest order through a projection onto basis functions keeping track of the no-slip condition and the continuity condition:
\begin{equation}
\label{eq1}\partial_x u+ \partial_y v +\partial_z w = 0.
\end{equation}
The experimentally observed coherence of velocity perturbations suggests us to take: 
$$
u = U_0 g_0(y) + U_1 g_1(y) + u',\qquad v = V_1 f_1(y) + v',\qquad w = W_0 g_0(y)+ W_1 g_1(y) + w',
$$
with~\cite{LM07,SM15}:
$$
f_1= A (1-y^2)^2,\qquad g_0= B (1-y^2), \qquad g_1=C y (1-y ^2),
$$
 where $A$, $B$, and $C$ are normalization constants.
 (Functions obtained {\it via\/} Empirical Orthogonal Decomposition of laboratory or computer data might preferably be used.)
Amplitudes $U_{0,1}$, $W_{0,1}$, and $V_1$ are functions of space $(x,z)$ and time $t$.
Quantities $u',w',v'$ are residues.
Equations governing les amplitudes are obtained by projecting the Navier--Stokes equations onto the three basis functions retained.
The continuity condition immediately yields:
$$
\partial_x U_0 + \partial_z W_0 = 0\qquad\mbox{and} \qquad \partial_x U_1+\partial_z W_1 =\beta V_1,
$$
where $\beta$ plays the role of a wave vector in the wall-normal direction ($\beta=\pi/2$ for stress-free boundary conditions and $\beta=\sqrt{3}$ in the no-slip case with the polynomials given above). 

Setting $Z_{0,1}=\partial_z U_{0,1} -\partial_x W_{0,1}$, one obtains~\cite{SM15}:
\begin{eqnarray}
\label{eq2} \nonumber &&\left(\uDelta -\beta^2\right)\partial_t V_1 = \nu\left( \uDelta^2 - 2 \beta^2 \uDelta + p_1\right) V_1 - \big(q \uDelta -\bar r \,\big) \big( \partial_x (U_0 V_1 ) +\partial_z ( W_0 V_1 )\big) \\
&& \hspace{1em}~+ r \left[\partial_{xx} \left( U_1 U_0 \right) + \partial_{xz}\left( U_1 W_0 + U_0 W_1 \right)+\partial_{zz} \left( W_0 W_1 \right)\right]+ \mathcal N_{V_1},\\
\label{eq3} \nonumber &&\partial_t  Z_0 
 +b\, \partial_x  Z_1 + \bar b\, \partial_z V_1 = \nu \left( \uDelta - \bar p_0\right)  Z_0 + s_0\left(\partial_{xz} \left( U_0 ^2 - W_0 ^2 \right) + (\partial_{zz}-\partial_{xx})( U_0 W_0 )\right] \\
 &&\hspace{1em}~+s_1\left[\partial_{xz} \left( U_1 ^2 - W_1 ^2 \right) + (\partial_{zz}-\partial_{xx})( U_1 W_1 )\right] -\bar s_0 \left[\partial_z( U_1 V_1 ) -\partial_x( W_1 V_1 ))\right] + \mathcal N_{Z_0},\\
\label{eq4} \nonumber &&\partial_t  Z_1 +b\, \partial _x  Z_0 = \nu \left( \uDelta - \bar p_1\right)  Z_1 +2s_1\left[\partial_{xz} ( U_1  U_0 - W_1 W_0 ) + (\partial_{zz}-\partial_{xx})( U_1 W_0 + U_0 W_1 )\right]\\
&&\hspace{1em}~ -\bar s_1 \left[\partial_z( U_0 V_1 ) -\partial_x( W_0 V_1 )\right] + \mathcal N_{Z_1},
\end{eqnarray}
where $\uDelta =\partial_{xx}+\partial_{zz}$ is the Laplacian in the plane of the flow.
Coefficients present in these equations are integrals of products of the basis functions and their derivatives over $y\in[-1,1]$ (see annex), the values of which are not interesting in themselves.
They can be obtained from~\cite{SM15} where the residues were also analyzed in view of going beyond the lowest significant order in a systematic and quantitative way.
In contrast, here we neglect all terms in $\mathcal N$ that account for the interaction between the retained amplitudes and the residues and between the residues themselves.
Accordingly, we get a system with a structure identical to that of the three-field model studied in~\cite[$\!$a]{LM07}, except for a velocity--vorticity treatment replacing the primitive-variable approach, which yields slightly different coefficients especially for the viscous terms (see also the remark at the end of the annex).
Numerical simulations of the Navier--Stokes equations have shown that this representation of the flow at first significant order contains about 90\% of the perturbation energy all along the range of Reynolds numbers corresponding to the transitional regime~\cite[$\!$Fig.$\,$B6]{Ma15}.
Here we shall be satisfied with this approximation though we recognize that, while the large-scale back-flows are reasonably well rendered at the qualitative level~\cite[$\!$b]{LM07}, numerical simulations of the so-truncated system show that the banded laminar--turbulent coexistence is not appropriately reproduced~\cite{Ma11} (which, by the way, suggested us to return to direct numerical simulations \cite{Ma1x}).

\section{Scale separation}

We are mostly interested in the problem of large scale flows empirically known to exist mostly from numerical simulations of Navier--Stokes equations~\cite{BT07,Ma1x,DS13} (see figure) but also detected in laboratory experiments \cite{CM15}.
According to the approach developed in~\cite[$\!$b]{LM07} these back-flows stem from source terms associated with the Reynolds stress tensor involving small-scale velocity fluctuations; the resulting system of equations is however not closed~\cite{BT07,DS13}.
Overcoming this limitation is the purpose of the present work, in which we proceed to an explicit separation of slow from fast space-time dependence of the velocity field.
To make this filtering concrete, in addition to the natural coordinates $x,z,t$ supposed to be rapidly varying we introduce slow coordinates $X,Z,T$. (Coordinate $y$ has been eliminated at the projection  step.)
In this multi-scale approach, noting $\mathcal Y$ the set of variables of interest $\{U_{0,1}; V_1; W_{0,1}\}$, we set:
\begin{equation}
\mathcal Y := \overline{\mathcal Y}(X,Z,T) + \widetilde {\mathcal Y}(X,Z,T|x,z,t).\label{eq5}
\end{equation}
The subcritical context does not allow us to implement the scale separation using some rigorous systematic perturbation expansion as can be performed in the supercritical case to account for the dynamics of convection cells or Taylor vortices close to their instability threshold~\cite{Ho06}.
So we assume a kind of absolute scale separation that leads us to consider that, when applied to the fields governed by (\ref{eq2},\ref{eq3},\ref{eq4}) the differential operators involving the natural coordinates are just splitted into operators involving the slow and fast coordinates separately.
In particular, the continuity condition (\ref{eq1}) yields:
\begin{eqnarray*}
\partial_x \widetilde U_0 + \partial_z \widetilde W_0 = 0\quad&\mbox{and}& \quad \partial_x\widetilde U_1+\partial_z \widetilde W_1 =\beta \widetilde V_1,\\
\partial_X \overline U_0 + \partial_Z \overline W_0 = 0\quad&\mbox{and}& \quad \partial_X\overline U_1+\partial_Z \overline W_1 =\beta \overline V_1.
\end{eqnarray*}

Let us  first consider the slowly varying large scales.
To obtain their governing equations, it suffices to substitute  (\ref{eq5}) into (\ref{eq2},\ref{eq3},\ref{eq4}) and just keep the terms involving the lowest order derivatives in the slow coordinates.
This yields:
\begin{eqnarray}
\label{eq8}
\beta^2\partial_T \overline V_1+ \underline{\nu p_1\overline V_1} &=& \mbox{}- \bar r \left(\underline{ \partial_X \overline{\widetilde U_0 \widetilde V_1}} + \partial_Z \overline{\widetilde W_0 \widetilde V_1}\,\right),\\
\label{eq9}\partial_T \overline Z_0 +\underline{ \nu\bar p_0 \overline Z_0} + \underline{ b \partial_X \overline Z_1} + \underline{\bar b \partial_Z \overline V_1}
&=& \bar s_0\left(\partial_X \overline{\widetilde W_1 \widetilde V_1}- \partial_Z \overline{\widetilde U_1 \widetilde V_1}\,\right),\\
\label{eq10}\partial_T \overline Z_1+ \underline{ \nu\bar p_1\overline Z_1} + \underline{ b \partial_X \overline Z_0} &=& \bar s_1 \left(\partial_X \overline{\widetilde W_0 \widetilde V_1} - \underline{ \partial_Z \overline{\widetilde U_0 \widetilde V_1}}\,\right),
\end{eqnarray}
in which overlined terms on the right hand side correspond to the slowly variable parts functions of the coordinates  $X,Z,T$.
Underlined terms are those retained in~\cite[$\!$b]{LM07} where the quasi-steady regime ($\partial_T\equiv 0$) was considered, while assuming that the other terms on the right hand side did not contribute for statistical symmetry reasons~\cite[$\!$a,$\,$\S2.4]{LM07}.
This assumption is valid in the absence of  ``$z\leftrightarrow -z$'' symmetry breaking but cannot be maintained in full generality since it is not verified in the late stage of spot growth~\cite{DS13} and {\it a fortiori\/} when an oblique laminar--turbulent pattern is present~\cite{Petal03}.

The problem is therefore to evaluate Reynolds stresses on the right hand side of (\ref{eq8}--\ref{eq10}) involving the average of products of small-scale  fast fluctuations. 
To estimate them, we notice that the small-scale flows involved are induced by the SSP that, following Waleffe~\cite{Wa97}, we approach within the framework of the MFU approximation ({\it local\/} view point as opposed to the {\it global\/} perspective related to large scales).
The eight-variable model that he introduced accounts for a simplified dependence of the flow in the $(x,z)$ plane, further Fourier analyzed in series truncated beyond their first-harmonic terms. 
The modes he chose correspond to a specific phase choice that insures strict spatial resonance between the different velocity  components involved.
Here we keep the MFU assumption considering a periodized domain $\ell_x\times 2\times\ell_z$ and the first harmonic approximation with basic wave vectors $\alpha=2\pi/\ell_x$ and $\gamma=2\pi/\ell_z$.
However we use a more general decomposition of the hydrodynamic fields permitting phase shifts between Fourier modes, which in a certain sense allows us to incorporate the effects of translational invariance in the local couplings~\cite{Ho06}.
Defining a velocity potential $\phi$ and a stream function $\psi$ from which the velocity component derive as  $u =\mbox{} -\partial_z \psi + \partial_x \phi$, $v=\mbox{}-\uDelta\, \phi$, $w=\partial_x \psi + \partial_z \phi$, so that  $\zeta=\mbox{}-\uDelta\, \psi$, we take the most general expressions possible:
 \begin{eqnarray*}
&& \Psi_0 = - {\overline U}_0 z + {\overline W}
_0 x + A_0 \cos\alpha x + \underline{B_0} \sin\alpha x + C_0 \cos\gamma z+ \underline{D_0} \sin\gamma z\\
&&\quad\mbox{} + E_0 \sin\alpha x\cos\gamma z+ \underline{F_0} \cos\alpha x\cos\gamma z
+ G_0 \cos\alpha x \sin\gamma z+ H_0 \sin\alpha x \sin\gamma z,\\
&&\Psi_1= - \underline{{\overline U}_1} z + {\overline W}
_1 x + \underline{A_1} \cos\alpha x + B_1 \sin\alpha x + C_1 \cos\gamma z+ D_1 \sin\gamma z\\
&&\quad\mbox{} + \underline{E_1} \sin\alpha x \cos\gamma z + F_1 \cos\alpha x \cos\gamma z
+ G_1 \cos\alpha x \sin\gamma z + H_1 \sin\alpha x \sin\gamma z,\\
&&\Phi_1 = A'_1 \cos\alpha x + B'_1 \sin\alpha x + \underline{C'_1} \cos\gamma z + D'_1 \sin\gamma z\\
&&\quad\mbox{} + E'_1 \sin\alpha x \cos\gamma z + F'_1 \cos\alpha x \cos\gamma z
+ \underline{G'_1} \cos\alpha x \sin\gamma z + H'_1 \sin\alpha x \sin\gamma z.
\end{eqnarray*}
Underlined variables are those introduced under a different name in Waleffe's original eight-equation model (Wa97 in the following), with `w' as a subscript.
These are ${\overline U}_1\equiv M_{\rm w} - 1$, $ D_0 \equiv - U_{\rm w}/\gamma$, $ B_0 \equiv A_{\rm w}/\alpha$, $ F_0 \equiv - B_{\rm w}$, $ A_1 \equiv - C_{\rm w}/\alpha $, $ E_1 \equiv D_{\rm w}$, $ C'_1 \equiv V_{\rm w}/\gamma$, and $ G'_1 \equiv E_{\rm w}$.
The set $\{ {\overline U}_0; {\overline W}_0 ; {\overline U}_1 ; {\overline W}_1\}:=\overline{\mathcal U}$, overlined on purpose, obviously correspond to the local values of the components of the large scale flows introduced previously. 
The remaining set, $\{ A_0 ; B_0 ;\dots; G'_1 ; H'_1 \}:= \mathcal A$, contain all the space-varying contributions that are periodic at the period of the MFU.
Accounting for the small-scale coherence, these amplitudes are supposed to be functions of the fast time $t$, and the slow coordinates $X$, $Z$ and $T$.
The system governing the dynamics at the MFU scale is obtained in the usual way by inserting the expressions of the velocity components that derive from $\Psi_{0,1}$ and $\Phi_1$ as indicated above in (\ref{eq2}--\ref{eq4}), and next separating the different trigonometric lines in $x$ and $z$.
A set of 28 equations for 28 unknowns is obtained, that possess all the features, lift-up, viscous dissipation, quadratic advection nonlinearities, expected from Navier--Stokes equations for wall-bounded shear flows within the MFU framework.
Formally, it reads:
\begin{eqnarray}
\label{eq11}\mbox{$\frac{\rm d}{{\rm d} t}$} \mathcal A + \mathcal L\, \mathcal A &=&\mathcal M \left(\overline{\mathcal U}\right) \mathcal A + \mathcal N (\mathcal A,\mathcal A),\\
\label{eq12}\mbox{$\frac{\rm d}{{\rm d} t}$} \overline{\mathcal U} + \mathcal L' \,\overline{\mathcal U} &=& \mathcal N' (\mathcal A,\mathcal A).
\end{eqnarray}
The format of the RNL proceedings~\cite{MaRNL} did not allow us to give the full expression of system (\ref{eq11},\ref{eq12}) here to be found in the annex.
As to subsystem (\ref{eq12}), the original article in French only presented the equations for ${\overline U}_0$ and ${\overline U}_1$ [equations (\ref{E1}) and (\ref{E3}) in the annex]:
\begin{eqnarray}
\label{eq13}\mbox{$\frac{\rm d}{{\rm d} t}$} {\overline U}_0 + \nu \bar p_0 {\overline U}_0 &=& \mbox{$\frac14$}\gamma \bar s_0[2\gamma^2( C'_1 D_1 - C_1 D'_1 ) + \kappa^2 ( E_1 H'_1 + F_1 G'_1 - E'_1 H_1 - F'_1 G_1 )]\\
\label{eq14}\mbox{$\frac{\rm d}{{\rm d} t}$} {\overline U}_1 + \nu \bar p_1 {\overline U}_1 &=& \mbox{$\frac14$}\gamma \bar s_1[2\gamma^2( C'_1 D_0 - C_0 D'_1 ) + \kappa^2 ( F'_1 G_0 + E'_1 H_0 - E_0 H'_1 - F_0 G'_1 )]
\end{eqnarray}
where $\kappa^2=\alpha^2+\gamma^2$.
The first equation has no equivalent in Wa97.
The second one closely corresponds to the first equation of his system~\cite[p. 891]{Wa97} for $M_{\rm w}=1+{\overline U}_1$, in the present notations:
$$
\mbox{$\frac{\rm d}{{\rm d} t}$} {\overline U}_1 + \nu \bar p_1 {\overline U}_1 = \mbox{$\frac14$}\gamma \bar s_1[2\gamma^2 C'_1 D_0 - \kappa^2 F_0 G'_1].
$$
As an example for subsystem (\ref{eq11}) we give here only the equation for $D_0$ [equation (\ref{E8}) in the annex] attached to the component of $u$ in $\cos(\gamma z)$ independent of $x$, i.e. the main ingredient of turbulent {\it streaks\/}, $-U_{\rm w}/\alpha$ in Wa97. It reads:
\begin{eqnarray}
\nonumber
\mbox{$\frac{\rm d}{{\rm d} t}$} D_0 + \nu\kappa_0^\gamma D_0 - \bar b\, C'_1 &=& s_0 \gamma \big( C_0 {\overline W}_0 + \mbox{$\frac12$}\alpha (B_0 F_0 - A_0 E_0)\big)\\
\nonumber&&\mbox{} + s_1 \big[ \gamma \big(C_1 {\overline W}_1 + \beta C'_1 {\overline U}_1\big) + \mbox{$\frac12$} \alpha\gamma \big((B_1 F_1 - A_1 E_1) + \beta^2 (A'_1 E'_1-B'_1 F'_1)\big)\\
\label{eq15}&&\hspace{4em}\mbox{}- \mbox{$\frac12$}\alpha^2\beta (A_1G'_1 + A'_1G_1 + B_1 H'_1 + B'_1 H_1) \big)\big],
\end{eqnarray}
where, in addition to already introduced coefficients, $\kappa_0^\gamma = \gamma^2 + \bar p_0$.
As detailed in the annex, equations in (\ref{eq11}) are rather complicated but, here, the viscous relaxation term and the lift-up term on the left hand side are easily identified ($C'_1$ correspond to $V_{\rm w}/\gamma$ in Wa97).
Of the three uniform amplitudes ${\overline W}_0$, ${\overline U}_1$, and ${\overline W}_1$ corresponding to the contribution of $\>\mathcal M \left(\overline{\mathcal U}\right) \mathcal A\>$ in (\ref{eq11}) for that equation, only ${\overline U}_1$, i.e. $M_{\rm w} -1$, is present in Wa97.
In our notations the equation for $U_{\rm w}$ is reduced to:
$$
\mbox{$\frac{\rm d}{{\rm d} t}$} D_0 + \nu\kappa_0^\gamma D_0 - \bar b\, C'_1 = \mbox{$\frac12$} \alpha\gamma s_0 B_0 F_0 - \mbox{$\frac12$} \alpha\gamma s_1 A_1 E_1 - \mbox{$\frac12$} \alpha^2\beta s_1 A_1G'_1 + \gamma\beta s_1 C'_1 {\overline U}_1.
$$

Let us notice that, in addition to Wa97 that contains 8 equations, system (\ref{eq11},\ref{eq12}) possess two other closed subsystems with 15 equations.
The first one is made of $\{$Wa97$\,;\,$AU$_0\,;\,{\overline U}_0\,;\,D_1 \}$, where AU$_0:=\{ A_0\, ; E_0 \,; B_1\, ; F_1\, ; H'_1\}$ and the second one of $\{$Wa97$\,;\,$AW$_0\,;\,{\overline W}_0\,;\,B'_1 \}$ where AW$_0:=\{C_0\, ; G_0 \, ; H_1\, ; D'_1\, ; F'_1\}$.
These systems are closed in the sense of any trajectory initiated in the corresponding subspaces lives in them indefinitely.
Each subsystem AU$_0$ and AW$_0$ is made of amplitudes respectively associated with the generators of infinitesimal translations along $x$ for AU$_0$ and $z$ for AW$_0$.
For example a solution translated by $\delta x$ along $x$ can be deduced from any solution of Wa97 provided that the perturbation is reduced to $ A_0 =\epsilon_x B_0 $, $ E_0 =-\epsilon_x F_0 $, $ B_1 =-\epsilon_x A_1 $, $ F_1 =\epsilon_x E_1 $, and $ H'_1 =- \epsilon_x G'_1 $, with $\epsilon_x = \alpha\delta_x$ and it is easily checked on (\ref{eq13}) reduced to $\mbox{$\frac{\rm d}{{\rm d} t}$} {\overline U}_0 + \nu \bar p_0 {\overline U}_0 = \mbox{$\frac14$}\gamma \bar s_0 \kappa^2 ( E_1 H'_1 + F_1 G'_1 )$, other terms being absent by assumption, that such a solution generates no uniform component $ {\overline U}_0$.
In contrast, any arbitrary perturbation in  AU$_0$ will generically induce some back-flow ${\overline U}_0$.
The same is true with perturbations in AW$_0$ and ${\overline W}_0$.
Back-flows $\overline{\mathcal U}$ are therefore naturally part of the dynamics of the complete 28-amplitude generalized Waleffe system.

Now, if we assume that this system only describes the small-scale (MFU) flow, and that the corresponding local solution can experience slow modulations governed by (\ref{eq8},\ref{eq9},\ref{eq10}), we see nonlocal back-flows appear {\it via\/} the terms in $\>\mathcal M \left(\overline{\mathcal U}\right) \mathcal A\>$, an example being given by equation (\ref{eq15}) with the presence of ${\overline W}_0$, ${\overline U}_1$ and ${\overline W}_1$ on its right hand side.
These terms account for the {\it advection\/} in a wide sense of the small scales of the flow, i.e. the amplitudes belonging to $\mathcal A$.

\section{Discussion}

In a purely phenomenological approach resting on a Ginzburg--Landau appropriate for a sub-critical bifurcation, Hayot and Pomeau~\cite{HP94} introduced a nonlocal feedback between the large scale recirculation and the intensity of local turbulence by an integral condition.
System (\ref{eq8},\ref{eq9},\ref{eq10}) can be understood as a differential counterpart of this condition.

Now, in a phase-dynamics perspective~\cite{Ho06}, essential in the context of extended systems of experimental interest, the introduction of two groups of additional amplitudes AU$_0$ and AW$_0$ associated with the infinitesimal translation generators in the plane of the flow bring a novel opening apt to raise the rigidity inherent in the MFU assumption, since the strictly fixed dimensions of the MFU become in some sense softened by the possibility of wavelength and orientation modulations {\it via\/} the analogue of Eckhaus and zig-zag instabilities in convection \cite{Ho06}, but here in the fully nonlinear regime far from any instability threshold.
As soon as the trivial contribution  ${\overline U}_1$ producing the S-shaped mean turbulent flow profile is no longer uniform~\cite{LM07,DS13}, spatiotemporally slowly varying back-flows ${\overline U}_0 $, ${\overline W}_0$ and ${\overline W}_1$ are unavoidably  generated. This observation leads us to think that an important factor --couplings {\it via\/} (\ref{eq8},\ref{eq9},\ref{eq10})-- has been pointed out relative to laminar--turbulent coexistence in transitional wall-bounded flows.
The argument is not limited to system (\ref{eq2},\ref{eq3},\ref{eq4}) where terms in $\mathcal N$ are neglected but, through the adiabatic elimination of residues, would work on any more general effective model able to render the oblique turbulent band regime in PCF and other wall-bounded plane shear flows. (It can be noticed that oblique bands were observed in numerical simulations of the two-dimensional model extending (\ref{eq2},\ref{eq3},\ref{eq4}) beyond first significant order~\cite{SM15}).

In a previous work~\cite{Ma12}, starting from a simplified four-equation Waleffe model, we developed a theory of the emergence of periodically modulated turbulent intensity.
In a purely phenomenological way, we allowed the diffusion of an inhomogenous distribution of the corresponding four amplitudes in an arbitrary direction of space and showed that the laminar--turbulent alternation could arise from a Turing mechanism when the diffusivity of the different amplitudes were sufficiently far apart.
We also called for the derivation of a model from the primitive equations, explaining such a turbulent diffusion.
The approach presented above is proposed as an element of answer along these lines but we also see that the result is not at all in the expected form:
The advection of small scales by large scale back-flows~\cite{DS13,Po15} appears to be much more natural than the diffusion induced by some anisotropic effective turbulent viscosity that does not seem to come out easily from the primitive equations.

\paragraph{Acknowledgements.} The author wants to express his gratitude to G. Kawahara and M. Shimizu (Osaka), T. Tsukahara (Tokyo), Y.~Duguet (LIMSI), R.~Monchaux and M.~Couliou (ENSTA-ParisTech) for numerous discussions, in particular within the framework of the {\sc jsps-cnrs} program {\sc transturb} ``Localized turbulent structures in transitional wall-bounded flows,'' as well as to K. Seshasyanan who took part in the development and numerical simulation of the model on which part of the discussion relies.

\newpage


\setcounter{equation}{0}

\noindent {\large \bf Annex: Generalized Waleffe model}\\[2ex]
For more readability, amplitudes appearing in {\CCB Wa97} are colored in {\it\CCB blue} (underlined in the main text);
the set {\CCR $\{$AU$_0$;${\overline U_0}\}$} is here painted  in {\it\CCR red}, and the set {\CCG $\{$AW$_0$;${\overline W_0}\}$} in {\it\CCG green}.
The guess to be introduced in equations (\ref{eq2},\ref{eq3},\ref{eq4}) is repeated here for convenience with the appropriate colors:
\BAN
\NN && \Psi_0 = - \UZ z + \WZ x +  \AZ \cos\alpha x + \BZ \sin\alpha x + \CZ \cos\gamma z+ \DZ \sin\gamma z\\
\NN &&\quad\mbox{} + \EZ \sin\alpha x\cos\gamma z+ \FZ \cos\alpha x\cos\gamma z+ \GZ \cos\alpha x \sin\gamma z+ \HZ \sin\alpha x \sin\gamma z,\\
&& \NN \Psi_1= - \UO z + \WO x + \AO \cos\alpha x + \BO \sin\alpha x + \CO \cos\gamma z+ \DO \sin\gamma z\\
\NN &&\quad\mbox{} + \EO \sin\alpha x \cos\gamma z + \FO \cos\alpha x \cos\gamma z
+ \GO \cos\alpha x \sin\gamma z + \HO \sin\alpha x \sin\gamma z,\\
\NN &&\Phi_1 = \AP \cos\alpha x + \BP \sin\alpha x + \CP \cos\gamma z + \DP \sin\gamma z\\
\NN &&\quad\mbox{} + \EP \sin\alpha x \cos\gamma z + \FP \cos\alpha x \cos\gamma z + \GP \cos\alpha x \sin\gamma z + \HP \sin\alpha x \sin\gamma z.
\EAN
The velocity and wall-normal vorticity components deriving from these fields read:
\BAN
U_0 = \UZ - \partial_z \widetilde \Psi_0,&& W_0 = \WZ + \partial_x \widetilde \Psi_0,\\
U_1 = \UO + \partial_x \Phi_1 - \partial_z \widetilde \Psi_1, &\qquad V_1=-\Delta\Phi_1,\qquad& W_1 = \WO + \partial_z \Phi_1 + \partial_x \widetilde \Psi_1,\\
Z_0=\partial_z U_0 -\partial_x W_0 = -\Delta \widetilde  \Psi_0,&&
\,\, Z_1=\partial_z U_1 -\partial_x W_1 = -\Delta \widetilde \Psi_1\,,
\EAN
where $\widetilde \psi_0$ and $\widetilde \Psi_1$ denote the periodically varying parts of $\Psi_0$ and $\Psi_1$ defined above.
All the coefficients introduced in model (\ref{eq2},\ref{eq3},\ref{eq4}) can be numerically evaluated once the shapes of the Galerkin basis functions are specified.
Values given below are obtained using the polynomials mentioned in the main text~\cite{SM15}.
Linear viscous coefficients and non-normal linear coupling constants read:%
\footnote{In the stress-free case one would get $\beta=\pi/2$, $\bar p_0 =0$, $\bar p_1 = \beta^2$, $p_1=\beta^4$, $b=1/\sqrt2$, and $\bar b = \pi/2\sqrt2$ \cite[$\!$a]{LM07}.}
$$
\begin{tabular}{l l l l l l}
$\bar p_0=\sfrac52$, &\quad $\bar p_1=\sfrac{21}2,$&\quad $\beta^2=3$,&\quad $p_1=\frac{63}2$, &\quad $b = \frac1{\sqrt{7}} \approx 0.3780,$ &\quad $\bar b = \sqrt{\frac{27}{28}}\approx 0.9820$,
 \end{tabular}
 $$
while constants relative to the nonlinear interactions are:\\
$$
\begin{tabular}{ l l l l}
$q = \frac{\sqrt{375}}{22}\approx0.8802$, &\quad $r = \frac{\sqrt{5}}4\approx0.5590$,&\quad $\bar r =-\frac{\sqrt{135}}{4}\approx -2.9047$,\\[1ex]
$s_0=\frac{\sqrt{135}}{14}\approx0.8299$, &\quad $ s_1=\frac{\sqrt{15}}{6}\approx0.6455$,&\quad
$ \bar s_0 = \frac{\sqrt{5}}{4}\approx0.5590$, &\quad $\bar s_1=-\frac{\sqrt{45}}{4}\approx-1.6771$.
 \end{tabular}
 $$
The model involves $\alpha=2\pi/\ell_x$ and $\gamma=2\pi/\ell_z$ as parameters, $\ell_x$ and $\ell_z$ being the in-plane dimensions of the MFU.
Accordingly,  in addition to the coefficients given previously, we introduce:
 $$
\begin{tabular}{ l l l l l l}
 $\kappa^2=\alpha^2+\gamma^2$, &\quad $\bar\alpha = \alpha/\kappa$, &\quad$\bar\gamma= \gamma/\kappa$, &\quad $\tau=\bar \gamma/\bar \alpha$,&\quad $g=\bar\alpha^2-\bar\gamma^2$, &\quad $g'=2\bar\alpha\bar\gamma$,
 \end{tabular}
 $$
 which will simplify the expressions of some coefficients of the nonlinear terms.
 Let $\theta$ be the angle between the streamwise direction and the in-plane diagonal of the MFU, then  $\tau=\tan\theta$, $g=\cos 2\theta$, and $g'=\sin2\theta$, relate to the shape of the MFU.

Subset (\ref{eq11}) of system (\ref{eq11},\ref{eq12}) relative to the periodically varying flow components contains three parts.
The first one, coming from the identification of amplitudes in $\Psi_0$, introduces constants
$$
\begin{tabular}{ l l l l l }
$\bar\kappa^2_\alpha=\alpha^2 +\bar p_0$, &\quad $\bar\kappa^2_\gamma = \gamma^2+\bar p_0$, &\quad $\bar\kappa^2_{\alpha\gamma} = \alpha^2+\gamma^2+\bar p_0$,&\quad
$ s_2 =  \bar s_0 + \beta s_1 $,&\quad $  s_3  = \bar s_0 + 2\beta  s_1$,
 \end{tabular}
$$
and reads:
\BA
\NN \DDt \AZ + \nu \bar\kappa^2_\alpha \AZ & = & - \alpha b \BO - s_0 \left( \alpha \BZ \UZ +\sfrac12 \alpha \gamma \left( \CZ \HZ - \DZ \EZ \right) \right)\\
\NN &&\mbox{}- s_1 \left(\alpha \BO \UO +\sfrac12\alpha\gamma\left ( \CO \HO - \DO \EO \right) + \sfrac12\beta\gamma^2 \left( \CO \FP + \DO \GP \right) \right)\\
&&\mbox{}+ s_2\left (\alpha \BP \WO +\sfrac12\alpha\beta\gamma\left ( \CP \HP - \DP \EP \right) - \sfrac12\gamma^2\left( \CP \FO + \DP \GO \right) \right),
\label{E5}\\
\NN \DDt \BZ + \nu\bar\kappa^2_\alpha \BZ & = &\alpha b \AO + s_0 \left (\alpha \AZ \UZ + \sfrac12\alpha\gamma\left( \CZ \GZ - \DZ \FZ \right) \right)\\
\NN &&\mbox{}+ s_1 \left( \alpha \AO \UO +\sfrac12\alpha\gamma\left( \CO \GO - \DO \FO \right) -\sfrac12\beta\gamma^2\left( \CO \EP + \DO \HP \right)\right)\\
&&\mbox{}- s_2\left( \alpha  \AP  \WO +\sfrac12\alpha\beta\gamma\left( \CP \GP - \DP \FP \right)+\sfrac12\gamma^2\left(\CP \EO + \DP \HO \right)\right),
\label{E6}
\EA
\BA
\NN \DDt \CZ + \nu\bar\kappa^2_\gamma \CZ & = & - \bar b \DP - s_0 \left( \gamma \DZ \WZ + \sfrac12\alpha\gamma\left( \BZ \GZ - \AZ \HZ \right)\right)\\
\NN &&\mbox{}- s_1 \left(\gamma \DO \WO +\beta\gamma \DP \UO + \sfrac12\alpha\gamma\left( \BO \GO - \AO \HO 
\right) + \sfrac12\alpha\beta^2\gamma \left( \AP  \HP - \BP \GP \right)\right.\\
&&\hspace{8ex}\left. \mbox{}+\sfrac12\alpha^2\beta \left( \AO \FP +  \AP  \FO + \BO \EP + \BP \EO \right)\right),
\label{E7}\\
\NN \DDt \DZ + \nu\bar\kappa^2_\gamma \DZ & = & \bar b \CP + s_0 \left(\gamma \CZ \WZ +\sfrac12\alpha\gamma\left( \BZ \FZ - \AZ \EZ \right)\right)\\
\NN &&\mbox{} + s_1 \left(\gamma \CO \WO +\beta\gamma \CP \UO + \sfrac12\alpha\gamma\left( \BO \FO - \AO \EO\right)+\sfrac12\alpha\beta^2\gamma \left( \AP  \EP - \BP \FP \right)\right.\\
 &&\left.\hspace{8ex} \mbox{}- \sfrac12\alpha^2\beta \left( \AO \GP + \AP \GO + \BO \HP + \BP \HO \right)\right),
\label{E8}\\
\NN \DDt \EZ + \nu\bar\kappa^2_{\alpha\gamma}  \EZ  &  =    & \alpha b  \FO      -  \gamma     \bar b   \HP +  s_0  \left(\alpha \FZ   \UZ  - \gamma  \HZ   \WZ -\alpha\gamma g  \AZ   \DZ \right)\\
\NN&&\mbox{}+  s_1  \left(\alpha  \FO   \UO  - \gamma  \HO  \WO -\beta\gamma  \HP   \UO -\alpha\gamma g  \AO   \DO  - \sfrac12{g'}^2 \kappa^2\beta  \BP   \CO \right)\\
 &&\mbox{} -\alpha  s_2  \FP  \WO  - \sfrac14{g'}^2 \kappa^2 s_3   \BO   \CP  + \alpha\beta\gamma(\bar\alpha^2 \bar s_0 + g \beta  s_1 )  \AP   \DP, 
\label{E9}\\
\NN\DDt  \FZ  + \nu \bar\kappa^2_{\alpha\gamma}  \FZ  &  =    & -\alpha b  \EO  -\gamma  \bar b   \GP - s_0  \left(\alpha \EZ   \UZ  + \gamma  \GZ   \WZ -\alpha\gamma g \BZ   \DZ \right)\\
\NN &&\mbox{}
-  s_1  \left(\alpha \EO   \UO  + \gamma  \GO  \WO + \beta\gamma  \GP   \UO 
-\alpha\gamma g  \BO  \DO +\sfrac12{g'}^2 \kappa^2 \beta  \CO   \AP \right)\\
&&\mbox{}
 +\alpha  s_2  \EP  \WO -  \sfrac14{g'}^2 \kappa^2  s_3   \AO   \CP 
 -\alpha\beta\gamma(\bar\alpha^2 \bar s_0 + g\beta  s_1 )  \BP   \DP ,
\label{E10}\\
\NN\DDt  \GZ  + \nu\bar\kappa^2_{\alpha\gamma}  \GZ  &  =    &  \gamma  \bar b   \FP  - \alpha b  \HO + s_0 \left(\gamma  \FZ   \WZ  - \alpha  \HZ   \UZ - \alpha\gamma g  \BZ   \CZ \right)\\
\NN &&\mbox{}
+  s_1  \left(\gamma  \FO   \WO -\alpha  \HO   \UO + \beta \gamma  \FP   \UO   - \alpha\gamma g  \BO   \CO   - \sfrac12{g'}^2\kappa^2 \beta   \DO   \AP \right) \\
&&\mbox{}  +\alpha  s_2  \HP  \WO -  \sfrac14{g'}^2 \kappa^2  s_3  \AO   \DP +\alpha\beta\gamma(\bar\alpha^2 \bar s_0 + g\beta  s_1 )  \BP   \CP ,
\label{E11}\\
\NN \DDt  \HZ  + \bar\kappa^2_{\alpha\gamma}  \HZ  &  =    & \alpha b  \GO  + \gamma  \bar b   \EP + s_0  \left(\gamma  \EZ   \WZ + \alpha  \GZ   \UZ  +\alpha\gamma g  \AZ   \CZ \right)\\
\NN &&\mbox{}+  s_1  \left(\gamma  \EO  \WO+ \alpha  \GO   \UO  +   \beta\gamma  \EP   \UO+ \alpha\gamma g  \AO   \CO  -   \sfrac12{g'}^2\kappa^2 \beta  \BP  \DO \right)\\
&&\mbox{}  -\alpha  s_2 \GP  \WO  -  \sfrac14{g'}^2 \kappa^2  s_3 \BO   \DP - \alpha\beta\gamma(\bar\alpha^2 \bar s_0 + g\beta  s_1 )  \AP   \CP .
\label{E12}
\EA
A second set stems from amplitudes involved in $\Psi_1$, introducing constants
$$
\begin{tabular}{ lll l }
$\kappa^2_\alpha=\alpha^2 +\bar p_1$, &\quad $\kappa^2_\gamma = \gamma^2+\bar p_1$, & \quad$\kappa^2_{\alpha\gamma} = \alpha^2+\gamma^2 +\bar p_1$, &\quad $ s_4 =     \bar s_1 + 2\beta  s_1 $.
\end{tabular}
$$
It reads:
\BA
\NN\DDt \AO + \nu \kappa^2_\alpha  \AO &   =    & - \alpha  b   \BZ +   s_1 \left(\alpha\gamma (  \DZ   \EO + \DO   \EZ -  \CZ   \HO - \CO   \HZ  ) - 2 \alpha(  \BZ   \UO +  \BO   \UZ )\right.\\
&&\hspace{4ex}\left.  \mbox{}-\beta\gamma^2 ( \CZ   \FP  +  \DZ   \GP )\right)+  s_4 \left(\alpha  \BP   \WZ  -\sfrac12 \gamma^2(\CP   \FZ + \DP   \GZ )\right),
\label{E13}\\
\NN \DDt  \BO  + \nu\kappa^2_\alpha   \BO &   =    & \alpha  b   \AZ  +  s_1  \left(\alpha\gamma(   \CZ   \GO + \CO   \GZ   -  \DZ   \FO  - \DO   \FZ ) + 2\alpha(  \AZ   \UO + \AO   \UZ ) \right.\\
&& \hspace{4ex} \left. \mbox{}-\beta\gamma^2 ( \CZ   \EP  +  \DZ   \HP )\right) -  s_4  \left(\alpha  \AP   \WZ  +\sfrac12\gamma^2(  \DP   \HZ  +  \CP   \EZ )\right),
\label{E14}\\
\NN \DDt  \CO  + \nu\kappa^2_\gamma  \CO  &  =    &  s_1  \left( \alpha\gamma (  \AZ   \HO + \AO   \HZ -  \BZ  \GO -  \BO   \GZ ) - 2\gamma(  \DZ   \WO +  \DO   \WZ ) \right.\\
&& \hspace{4ex}\left.  \mbox{} -  \alpha^2\beta (  \AZ   \FP + \BZ  \EP )\right) -  s_4  \left(\gamma  \DP   \UZ +\sfrac12\alpha^2( \AP   \FZ + \BP   \EZ ) \right),
\label{E15}\\
\NN \DDt  \DO  + \nu \kappa^2_\gamma  \DO  &  =    & s_1  \left(\alpha\gamma(  \BZ   \FO + \BO   \FZ   -  \AZ   \EO- \AO   \EZ ) + 2\gamma( \CZ   \WO +  \CO   \WZ )\right. \\
&& \hspace{4ex}\left. \mbox{} - \alpha^2\beta ( \AZ   \GP + \BZ   \HP)\right) +  s_4  \left(\gamma  \CP   \UZ  - \sfrac12\alpha^2( \AP   \GZ  +    \BP \HZ )\right),
\label{E16}\\
\NN\DDt  \EO  + \nu\kappa^2_{\alpha\gamma}  \EO  &  =    & \alpha  b   \FZ 
+  s_1  \left( 2\bar\alpha^2 \alpha (   \FZ   \UO+ \FO   \UZ) - 2\bar\gamma^2 \gamma(  \HZ  \WO+ \HO   \WZ) \right.\\
\NN && \hspace{4ex} \left. \mbox{}-2\alpha\gamma g(  \AZ   \DO + \AO   \DZ )\vphantom{ \HZ  \WO } \right) -  \sfrac14{g'}^2\kappa^2  s_4  \left( \BZ   \CP  +  \CZ   \BP \right)\\
&&\hspace{4ex} \mbox{}  - \alpha(\bar s_1 + 2 \bar\alpha^2 \beta  s_1 )  \FP   \WZ  - \gamma(\bar s_1 + 2\bar\gamma^2 \beta  s_1 ) \HP   \UZ ,
\label{E17}\\
\NN \DDt  \FO  + \nu\kappa^2_{\alpha\gamma}  \FO  &  =    & - \alpha  b   \EZ 
-  s_1  \left(2\bar\alpha^2\alpha (  \EZ   \UO + \EO   \UZ )+ 2 \bar\gamma^2 \gamma(  \GZ  \WO + \GO   \WZ ) \right.\\
\NN && \hspace{4ex} \left.  \mbox{}+  2\alpha\gamma g( \BZ   \DO + \BO   \DZ )\vphantom{ \HZ  \WO } \right) -  \sfrac14{g'}^2\kappa^2  s_4  (  \AZ   \CP +  \AP  \CZ ) \\
&&\hspace{4ex} \mbox{}+\alpha(\bar s_1 + 2 \bar\alpha^2 \beta  s_1 )  \EP   \WZ  - \gamma(\bar s_1 + 2\bar\gamma^2 \beta  s_1 )  \GP   \UZ ,
\label{E18}
\EA
\BA
\NN\DDt  \GO  + \nu\kappa^2_{\alpha\gamma}  \GO  &  =    & - \alpha  b   \HZ 
-  s_1  \left( 2\bar\alpha^2\alpha( \HO   \UZ  +  \HZ   \UO ) - 2\bar\gamma^2 \gamma\left( \FO   \WZ  +  \FZ  \WO \right)\right.\\
\NN && \hspace{4ex} \left. \mbox{}+ 2\alpha\gamma g( \BZ   \CO  +  \BO   \CZ )\vphantom{ \HZ  \WO }\right)- \sfrac14{g'}^2\kappa^2  s_4   \left(  \AZ   \DP  +  \AP  \DZ \right)\\
&&\hspace{4ex}\mbox{}
 + \gamma(\bar s_1 + 2\bar\gamma^2 \beta  s_1 )  \FP   \UZ + \alpha(\bar s_1 + 2 \bar\alpha^2 \beta  s_1 )  \HP   \WZ ,
\label{E19}\\
\NN \DDt  \HO  + \nu\kappa^2_{\alpha\gamma}   \HO  &  =    & \alpha  b   \GZ 
+ s_1 \left(2\bar\alpha^2\alpha (  \GZ   \UO + \GO   \UZ )
+ 2\bar\gamma^2 \gamma ( \EZ   \WO  +  \EO   \WZ )\right.\\
\NN && \hspace{4ex} \left.\mbox{}+ 2\alpha\gamma g( \AZ   \CO  +  \AO   \CZ )\vphantom{ \HZ  \WO }\right)- \sfrac14{g'}^2\kappa^2  s_4  (  \BZ   \DP +  \BP  \DZ)\\
&& \hspace{4ex}\mbox{}
 -\alpha(\bar s_1 + 2 \bar\alpha^2 \beta  s_1 )  \GP   \WZ + \gamma(\bar s_1 + 2\bar\gamma^2 \beta  s_1 )  \EP   \UZ .
\label{E20}
\EA
The third and last set governs amplitudes involved in $\Phi_1$ and introduces:%
\footnote{In the stress-free for which $p_1=\beta^4$, one gets $\kappa'^4_\alpha=\mu^4_\alpha$,  $\kappa'^4_\gamma=\mu^4_\gamma$,  $\kappa'^4_{\alpha\gamma}=\mu^4_{\alpha\gamma}$, see [5].}
$$
\begin{tabular}{ lll }
$\mu^2_\alpha = {\alpha^2 + \beta^2}$, &\quad $\mu^2_\gamma = {\gamma^2 + \beta^2}$,&\quad $\mu^2_{\alpha\gamma} = \alpha^2+\gamma^2 + \beta^2$, \\[1ex]
$\kappa'^4_\alpha = \alpha^4 +2 \beta^2 \alpha^2 + p_1 $,&\quad $\kappa'^4_\gamma = \gamma^4 +2 \beta^2 \gamma^2 + p_1 $,&\quad $\kappa'^4_{\alpha\gamma} =\kappa^4 + 2 \beta^2 \kappa^2 + p_1$.\\[1ex]
$c_\alpha  =    \alpha( \bar r  + 2 \beta r    +\alpha^2 q  )$,&\quad $d_\alpha  =    \sfrac12 \gamma^2 \tau ( \bar r  + \alpha^2 q  )$,&\quad $e_\alpha  =     \sfrac12 \tau(\kappa^2 (\alpha^2 q  + \bar r ) + 2\alpha^2\beta r   )$,\\[1ex]
 $c_\gamma  =    \gamma( \bar r  + 2 \beta r    +\gamma^2 q  )$,&\quad $d_\gamma  =    \sfrac12\alpha^2\tau^{-1}(  \bar r +\gamma^2 q  )$,&\quad $e_\gamma  =     \sfrac12 \tau^{-1}(\kappa^2 (\gamma^2 q  + \bar r ) + 2\gamma^2\beta r   )$,\\[1ex]
$c_\kappa  =     ( \bar r  + 2\beta r    +\kappa^2 q  )$.
\end{tabular}
$$
It reads:
\BA
\NN &&\mu^2_\alpha\DDt  \AP  + \nu \kappa'^4_\alpha \AP   =    
\gamma^2 r   ( \CZ   \FO  +  \CO   \FZ +  \DZ   \GO  + \DO   \GZ  ) \\
 &&\hspace{4em}\mbox{}+ d_{\alpha} ( \CP   \HZ  -  \DP   \EZ )
  +e_\alpha ( \DZ   \EP  -  \CZ   \HP ) - c_\alpha \BP   \UZ  ,
\label{E21}\\
\NN && \mu^2_\alpha \DDt  \BP  + \nu \kappa'^4_\alpha  \BP  =    \gamma^2 r    (\CZ   \EO +  \CO   \EZ  +  \DZ   \HO   + \DO   \HZ) \\
&&\hspace{4em}\mbox{}+d_\alpha( \DP   \FZ - {\CCB{C}'_1}  \GZ )+e_\alpha(  \CZ   \GP  - \DZ   \FP ) + c_\alpha  \AP   \UZ ,
\label{E22}\\
\NN&&\mu^2_\gamma \DDt   \CP  + \nu\kappa'^4_\gamma  \CP   =    \alpha^2 r   (   \AZ   \FO  + \AO   \FZ +  \BZ   \EO + \BO   \EZ  )\\ 
&&\hspace{4em}\mbox{}+d_\gamma \left( \BP   \GZ  -  \AP   \HZ \right) +  e_\gamma( \AZ   \HP  -  \BZ   \GP )- c_\gamma  \DP   \WZ ,
\label{E23}\\
\NN&& \mu^2_\gamma\DDt  \DP  + \nu\kappa'^4_\gamma  \DP   =    \alpha^2 r    (   \AZ   \GO  +  \AO   \GZ +  \BZ   \HO +  \BO   \HZ )\\
 &&\hspace{4em}\mbox{}+ d_\gamma \left( \AP   \EZ  -  \BP   \FZ \right)+ e_\gamma ( \BZ   \FP  -  \AZ   \EP ) +  c_\gamma  \CP   \WZ ,
\label{E24}\\
\NN&&\mu^2_{\alpha\gamma} \DDt  \EP + \nu\kappa'^4_{\alpha\gamma} \EP   =      \sfrac12 g'^2 \kappa^2 r    (   \BZ   \CO  +  \BO   \CZ )\\
 &&\hspace{4em}\mbox{}  + c_\kappa\left((\alpha  \FP   \UZ -\gamma  \HP   \WZ ) +   \sfrac12{g'} (\gamma^2 \AZ   \DP - \alpha^2  \AP   \DZ )\right),
\label{E25}\\
\NN && \mu^2_{\alpha\gamma} \DDt  \FP  +\nu\kappa'^4_{\alpha\gamma}  \FP  =        \sfrac12 g'^2 \kappa^2 r    ( \AZ   \CO + \AO   \CZ ) \\  
&&\hspace{4em}\mbox{} -  c_\kappa\left((\alpha   \EP   \UZ  +\gamma  \GP   \WZ ) +   \sfrac12{g'} (\gamma^2 \BZ   \DP   -\alpha^2   \BP   \DZ ) \right),
\label{E26}\\
\NN && \mu^2_{\alpha\gamma}  \DDt  \GP  + \nu \kappa'^4_{\alpha\gamma}  \GP  =      \sfrac12 g'^2 \kappa^2 r    (  \AZ  \DO +  \AO   \DZ )\\
&&\hspace{4em}\mbox{} + c_\kappa \left((\gamma \FP   \WZ -\alpha  \HP   \UZ )+   \sfrac12{g'}(\gamma^2 \BZ   \CP -\alpha^2 \BP   \CZ )\right),
\label{E27}\\
\NN && \mu^2_{\alpha\gamma}   \DDt  \HP +\nu\kappa'^4_{\alpha\gamma}  \HP  =    \sfrac12 g'^2 \kappa^2 r    ( \BZ   \DO + \BO   \DZ )\\
 &&\hspace{4em}\mbox{} +  c_\kappa \left((  \gamma  \EP  \WZ  + \alpha  \GP   \UZ )      +   \sfrac12{g'}( \alpha^2  \AP   \CZ  - \gamma^2  \AZ   \CP )\right).
\label{E28}
\EA
Finally, the equations governing $\overline{\mathcal U}$, subset (10) of system (9,10) in the  main text explicitly read:
\BA
\DDt  \UZ + \nu \bar p_0  \UZ = \sfrac14\gamma \bar s_0[2\gamma^2(  \CP \DO - \CO \DP )
+ \kappa^2 (   \EO \HP  +  \FO \GP -  \EP \HO - \FP \GO )],
\label{E1}\\
\DDt \WZ +\nu \bar p_0  \WZ = \sfrac14 \alpha \bar s_0 [ 2\alpha^2( \AO  \BP - \AP \BO )+ \kappa^2 (  \EP  \FO  + \GO  \HP  - \EO \FP - \GP \HO)],
\label{E2}\\
\DDt  \UO + \nu \bar p_1  \UO = \sfrac14\gamma \bar s_1[2\gamma^2( \CP \DZ - \CZ \DP )  +  \kappa^2 ( \FP \GZ + \EP \HZ  - \EZ \HP  - \FZ \GP )],
\label{E3}\\
\DDt {\overline W}_1 + \nu \bar p_1 {\overline W}_1 = \sfrac14\alpha \bar s_1 (2\alpha^2 ( \AZ \BP - \AP \BZ ) + \kappa^2 ( \EP \FZ + \GZ  \HP - \EZ \FP - \GP \HZ )].
\label{E4}
\EA

Checking that energy is conserved by quadratic terms is not as easy as for previous Galerkin models, due to the uses of the velocity--vorticity formulation that eliminates pressure beforehand.
Galerkin projection over the chosen basis indeed does not commute with taking the curl of the primitive equations that involves taking derivatives with respect to the $y$ coordinate.
(With trigonometric lines appropriate to stress-free boundary conditions at the plates, it does.)
In the present approach, energy conservation is hidden in the complicated  expressions of the nonlinear interaction coefficients and coefficients $\mu$ in factor of the time derivatives in the equations arising from $\Phi_1$.
It could have been a little more conspicuous by making the following changes $(\AP,\BP) \mapsto\mu_{\alpha}(\AP,\BP)$, $(\CP,\DP) \mapsto \mu_\gamma(\CP,\DP)$, and $(\EP,\FP,\GP,\HP) \mapsto \mu_{\alpha\gamma} (\EP,\FP,\GP,\HP)$, while complicating further the formal expressions of the coefficients.
As a matter of fact, similar changes were also performed by Waleffe to obtain the final expression of {\CCB Wa87} in~\cite{Wa97}.

When restricted to the amplitudes involved in  {\CCB Wa97}, our system appears to be more generic than Waleffe's model owing to accidental cancellation implied by trigonometric relations in the stress-free case.
As a matter of fact, in equation (\ref{E13}) for $\AO$, a term $\FZ \CP$ is present that has no equivalent in {\CCB Wa97}: There is no term in $B_{\rm w}V_{\rm w}$ in equation (10.6) for $C_{\rm w}$, which invalidates the statement in~\cite{Wa97}, p.~892, middle of left column, about feedbacks 
among the triad $\{\FZ,\CP,\AO\}$ ($\equiv\{B_{\rm w},V_{\rm w},C_{\rm w}\})$.
(The detailed consequences of this observation have however not been scrutinized, nor the stability of putative fixed points of {\CCB Wa97}.)

Translational invariance properties can be checked by careful inspection of the equations set. For example, it is observed that an infinitesimal $x$-translation not only generates no $\UZ$ as stated in the main text, but also that the equations for the amplitudes in  {\CCR AU$_0$} are, at first order in $\epsilon_x=\alpha \delta x$, just proportional to the equations for the corresponding amplitudes in the parent solution of {\CCB Wa97} with the same proportionality coefficient $\epsilon_x$ and the appropriate signs.

Finally, when studying the stability of any nontrivial solution to {\CCB Wa97}, one cans split the tangent space into four subspaces.
The first one corresponds to {\it amplitude\/} perturbations and involves only perturbations within {\CCB Wa97}.
The second and third ones are respectively related to  $x$ and $z$ {\it phase\/} perturbations, namely $\{ ${\CCR AU$_0$}\,;\,{\CCR$\overline U_0$}\,;$\,D_1 \}$ and  $\{${\CCG AW$_0$}\,;\,{\CCG $\overline W_0$}\,;$\,B'_1 \}$.
We know from the translational invariance property that they possess a neutral mode that is isolated as long as one considers the low-dimensional system obtained by sticking to the MFU assumption, but at the origin of continuous phase branches when the assumption is relaxed~\cite{Ho06}. (The presence of $D_1$ and $B'_1$ in these groups is easily shown not to break the neutrality of the phase perturbations with ${\CCR\overline U_0}=0$ and ${\CCG\overline W_0}=0$.)
The last group, $\{\WO; \HZ; \CO; \GO; \AP; \EP \}$, does not seem to be related to any specific symmetry.

Going beyond the remarks in \S4 and above is left for future work, in particular the search for explicit nontrivial solutions to {\CCB Wa97} with its no-slip coefficients and their general stability properties.

\end{document}